\documentclass[9pt]{article}
\usepackage{latexsym}
\usepackage{graphicx}
\usepackage{amsfonts,amssymb}

\begin{document}
\title{\textbf{Orientation in Social Networks}}
\author{Yanqing Hu\footnote{yanqing.hu.sc@gmail.com}, Ying Fan, Zengru Di\footnote{zdi@bnu.edu.cn}\\
\\\emph{Department of Systems Science, School of Management,}\\
\emph{Center for Complexity Research,}
\\\emph{Beijing Normal University,
Beijing 100875, P.R.China}} \maketitle
\begin{abstract}

Stanley Milgram's small world experiment presents ``six degrees of
separation" of our world. One phenomenon of the experiment still
puzzling us is that how individuals operating with the social
network information with their characteristics can be very adept at
finding the short chains. The previous works on this issue focus
whether on the methods of navigation in a given network structure,
or on the effects of additional information to the searching
process. In this paper, we emphasize that the growth and shape of
network architecture is tightly related to the individuals'
attributes. We introduce a method to reconstruct nodes' intimacy
degree based on local interaction. Then we provide an intimacy based
approach for orientation in networks. We find that the basic reason
of efficient search in social networks is that the degree of
``intimacy" of each pair of nodes decays with the length of their
shortest path exponentially. Meanwhile, the model can explain the
hubs limitation which was observed in real-world experiment.

\end{abstract}

China has a famous shortest poem titled \emph{life: net} (by poet
Bei Dao). This one word poem tells us that any one in the world is
living in the invisible social networks. In the 1960s, Stanley
Milgram has revealed a striking feature of these networks by his
famous experiment \cite{the oldest experiment}. He showed us that
any two people can be linked by a short chain of friends. The
average length of the chains is about six, which is quite remarkably
close to Karinthys prediction 40 years earlier \cite{Earlier}. This
fascinating result has been popularized in the 1990s by John Guare's
successful play ``six degrees of separation" \cite{Play} and has
been known as small world phenomena. More recent empirical studies
using the Internet have demonstrate the similar conclusion \cite{new
experiment,News}.

Actually, Milgram's experiment showed us two issues of special
interest: first, is the existence of short paths, and second, is the
ability of people at finding them efficiently. The first issue has
been extensively studied, especially with the Watts-Strogatz
Small-World Network (SW) model (see Ref. \cite{first issue b,
SWNature} and references therein for review, such as \cite{BA
model,Newman-review}). Here we focus only on the second issue that
we still lack an equally complete understanding.

Kleinberg first noticed the searching problem and presented
excellent models with perfect mathematical analysis \cite{navigation
brief nature,navigation full}. He modeled the social network as a
lattice based network with some long-range connections. Under the
condition that every node knows the lattice coordinates of his
immediate neighbors (acquaintances) and the target, the letter
delivering process was: each letter holder (node) forwards the
letter across a connection that brings it as close as possible to
the target in lattice distance. Kleinberg found that the network is
searchable when the long-range connections obey a special
distribution. Moreover, he has given some results on hierarchical
network models \cite{Kleinberg hierarchical model}. Watts \textit{et
al} presented a hierarchical network based model which is more
approximate to the real-world social network and got some
interesting results by numerical simulation \cite{full model
navigation}. There also exist some other models try to present a
framework of the second issue such as combining random walks and
targeting searches at nodes with high degree \cite{power-law
networks search}. Greedy routing and its modification have been
studied extensively, by computer and social scientists \cite{Use
Kleinberg search,Fraigniauda,Fraigniaudb,Oskar licentiate
thesis,Analyzing kleinberg,one-d-Analyzing kleinberg}.

The above studies on network navigation have indeed captured some
basic features of Milgram's experiment, such as the letter
delivering process is mainly based on geographic proximity and
similarity of profession. But they all need some special network
structures. So can we develop an approach to navigate networks
efficiently without any requirement for network structure?

From the original Milgran's experiment to modern empirical studies
\cite{the oldest experiment}, we know that the letter holder chose
his next recipient based on their location, profession, education,
and other interests. It seems that without the information in
addition to the network structure, it is impossible to search
networks efficiently. So some researchers have suggested to overlap
another network that describe the relationships of individuals'
attributes such as location or profession to social network, so that
the social network is searchable when we combine the information of
these two networks \cite{power-law networks Kleinberg search}. But
we argue that the alleged additional information to networks is
indeed tightly related to social networks. Obviously, the
probability of acquaintance is actually related to the proximity
between individuals' attributes. The structural properties of social
network should be shaped by these factors. The formation and
evolution of social networks are affected or even determined by the
individual characters. That is why the individuals' attributes could
give us some information about network structures and the social
networks can be searched efficiently based on these factors.
Recently, just when we prepared this manuscript,
Bogu$\tilde{n}\acute{a}$ \emph{at al} have also indicated that
social distances among individuals have a role in shaping the
network architecture and that, at the same time, these distances can
be used to navigate the network. They discussed the effects of
hidden metric space to the node similarity and navigability of
networks \cite{scale free metric navigation}.

From above arguments, we know that the individuals' attitudes and
network structures are correlated each other tightly. The
individuals' attitudes affect the network evolution and thus they
should embedded in the network topology. Then, the problem becomes:
can we recover the embedded information about the nodes in networks
and use it to realize the effective searching?

In this article, we introduce a method to get nodes' intimacy
through local interactions. Then we present an approach of
orientation in network based on the ``intimacy degree". We assign an
$n$-dimensional vector to each node to describe its attribute. This
vector could have the information of other nodes in the network
through a series of acquaintance. It could be abstracted from the
network structure and can measure the ``intimacy degree" of a node
with any other nodes. A pair of nodes will be more intimacy if they
are near and less intimacy if they are far away in the network. The
process of delivering letter in the network is that the current
letter holder always forwards the letter to the candidate who has
the largest intimacy with the target. It has been demonstrated in
the following discussion that our approach is very efficient and can
be used in many network searching problems.

\section{Intimacy degree}
\begin{figure}
\center
\includegraphics[width=8cm]{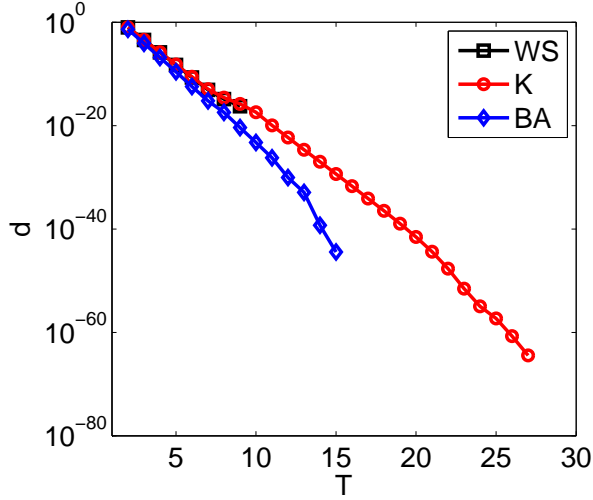}
\caption{Convergence speed. $d$ signifies the average value of all
the absolute differences of the intimacy degree between $T$ and
$T-1$ steps. The numerical experiments are done in WS (each node
link it's two nearest neighbors and one random long rang
connection), K (Kleinberg one dimensional small world networks, each
node link it's two nearest neighbors and one long rang connection
with clustering exponent $\alpha=1$), and BA (scale-free network
model with average degree 3) networks respectively. All the network
size is $n=1000$. From the plot we can see that the $d$ drop with
$T$ exponentially.}\label{Intensity decay soon}
\end{figure}

\begin{figure}
\center
\includegraphics[width=7cm]{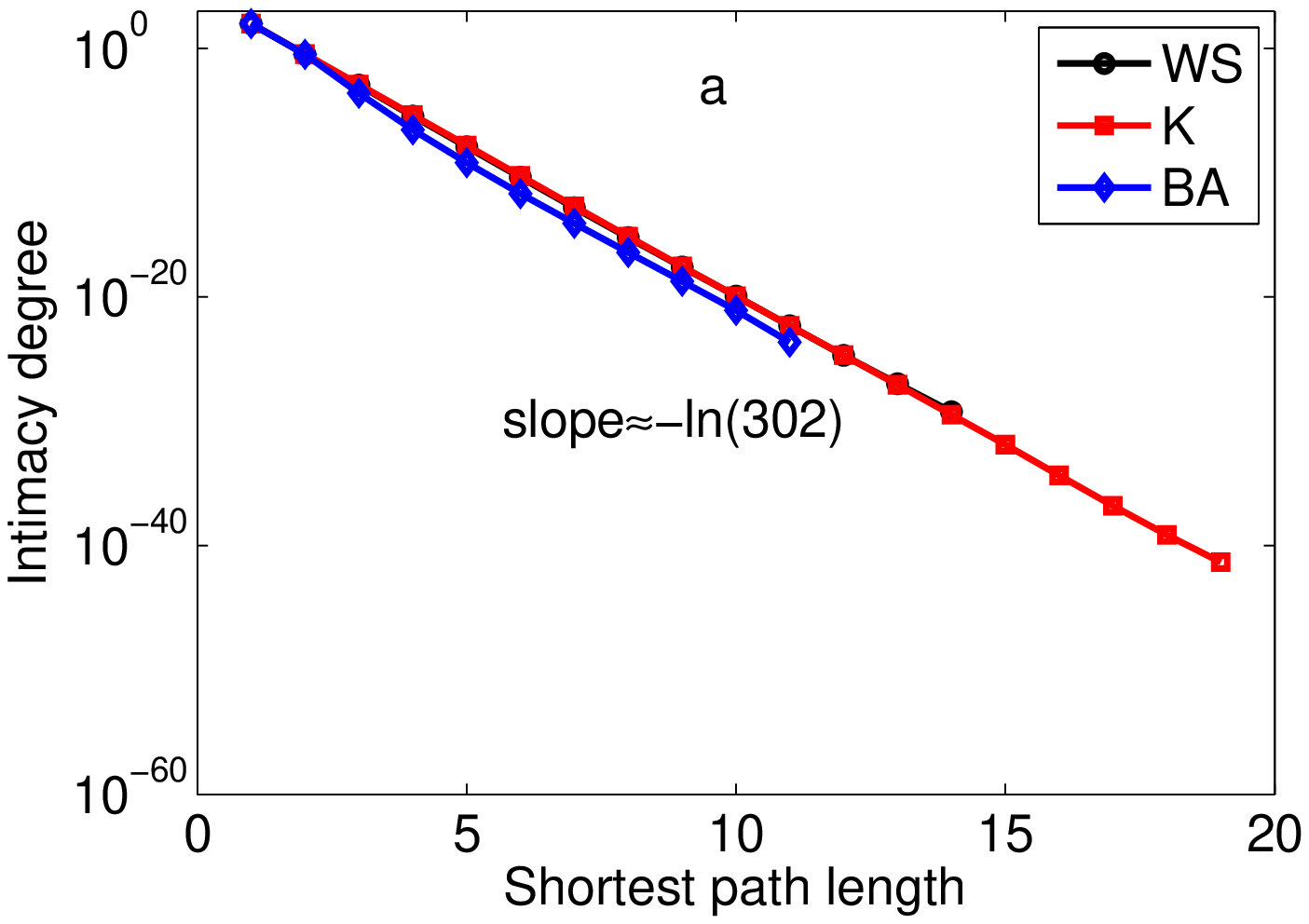}\includegraphics[width=7cm]{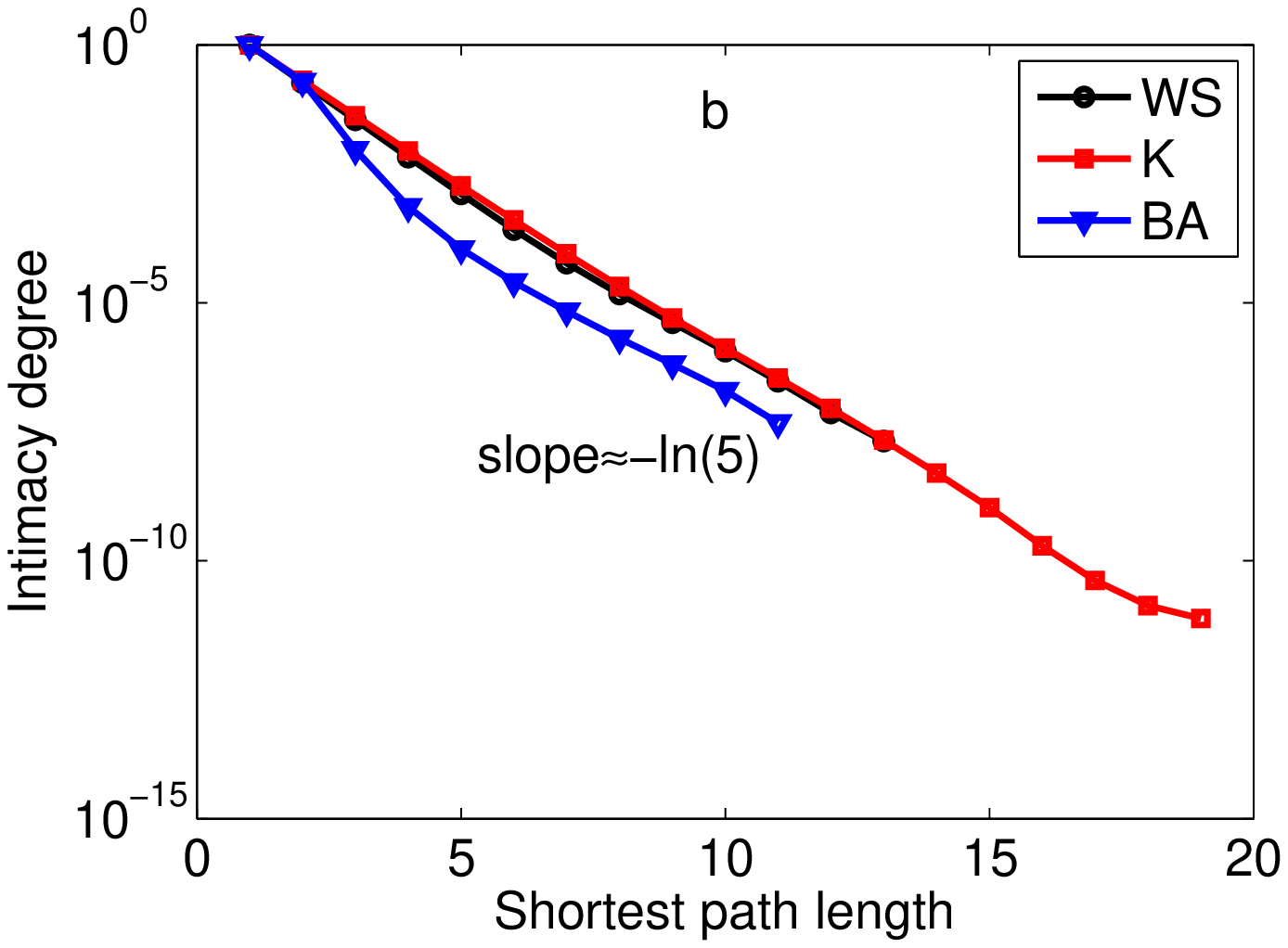}
\caption{Average Intimacy degrees decline exponentially with the
shortest path length. The simulations are done in the same networks
as in Fig.1. (A)$a=100$, the intimacy vectors evolved $100$ times.
(B) $a=1$, the intimacy vectors evolved $10$ times. Each point
denotes the average intimacy degree and the corresponding length of
the shortest path. From the two plots, we can safely conclude that
the intimacy degree of each pair of nodes decays with the length of
their shortest path exponentially. The slope approximate $-lnH$,
where $H=a(k+1)-1$ theoretically (See Supplementary Theorem 1). In
plot ($a$), $H=302$ and in plot ($b$), $H=5$. They are consistent
well with simulations.} \label{decay}
\end{figure}

Now we will reconstruct the individual's attributes related with the
network structure. Based on the assumption that social network
topology contains the individuals' attributes that affect the
network evolution. Here, an individual's attributes on the network
could described as its intimacy degree with other nodes. The
intimacy can be regarded as the integration of the similarities of
the occupations, hobbies, locations, or nationalities etc. Suppose
there is a connected network with $n$ nodes. We assign an
$n$-dimensional vector $\textbf{v}_{i}$ to each node $i$. Its $j$th
element $v_{i}(j)$ denotes the degree of intimacy of node $i$ to
node $j$. If $i=j$, we set $v_{i}(j)=a$ $(a\geq1)$, which indicates
the intimacy degree of each node to itself is a constant $a$ always.
In the initial, for each $i$ and $j$, we set $v_{i}(j)=a$ when
$i=j$, otherwise, $v_{i}(j)=0$. Each time we update the intimacy
vectors of every node by local interaction parallelly. Suppose node
$i$ has $k$ neighbors (in this paper, we say node $h$ is a neighbor
of node $i$ always means that there is an edge form node $i$ to node
$h$, if the network is a directed network), which are $N_{i}^{1},
N_{i}^{2}, \cdots, N_{i}^{k}$. $\textbf{v}_{N_{i}^{1}},
\textbf{v}_{N_{i}^{2}},\cdots,\textbf{v}_{N_{i}^{k}}$ are $k$
intimacy vectors of the $k$ neighbors. Then $\textbf{v}_{i}$ can be
updated through the interaction with its neighbors according to the
following four steps:
\begin{equation}(1)\textbf{v}_{i}=\sum_{h=1}^{k}\textbf{v}_{N_{i}^{h}};\ \
(2)v_{i}(i)=0;\ \
(3)\textbf{v}_{i}=\frac{\textbf{v}_{i}}{\sum_{j=1}^{n} v_{i}(j)};\ \
(4)v_{i}(i)=a; \label{evolution}\end{equation} In a word, we always
keep $v_{i}(i)=a$, and the sum of all the other elements of
$\textbf{v}_{i}$ is $1$. The intimacy vectors will be fixed within
$O(\ln n)$ steps of evolution (Fig.\ref{Intensity decay soon}). So
by a proper steps of iteration, the intimacy vector of every node
can be given. The element $v_{i}(j),(j\neq i)$ denotes the
comparative intimacy of node $i$ to node $j$. Our crucial findings
are that the degree of intimacy of each pair of nodes decays with
the length of their shortest path exponentially in statistical sense
(Sup. Theorem $1$ and Fig.\ref{decay}). It indicates that the degree
of intimacy of each pair nodes is dominated by the shortest path
between them in statistical sense. Does it contravene to common
sense? Suppose people only know $\frac{1}{w}$ of his neighbor's
information, where $w>1$. Then one will know only $\frac {1}{w^{2}}$
of information about his neighbor's neighbor. In this way we can
easy conclude that degree of intimacy of each pair of nodes (here we
regard the amount of information one knows about the other as
intimacy degree) decays with the length of their shortest path
exponentially.

\section{Orientation}
Obviously, in Milgram's experiment, current message holder always
try to forward the message to a immediate neighbor who seems can
send the message to the target most quickly. How to chose the
suitable neighbor? Suppose every one has only the local information
, that means each node knows and only knows his neighbor's intimacy
vectors, We may think that people will always send the message to
the neighbor who has the largest intimacy degree with the target. So
the orientation is the process that the current message holder $i$
sends the message to its neighbor $h$, which has the most intimacy
degree with the target $t$. And then node $h$ will send the message
in the same way until it reaches the target $t$. We can strictly
prove that the degree of intimacy will decline inversely to the
degree of the node by which intimacy passes (Sup. Theorem $1$). It
implies that ``highly connected individuals (hubs) appears to have
limited relevance to the kind of social search" which was observed
in real-world experiment \cite{new experiment}. Moreover we also can
prove that for a connected network, the chosen neighbor $h$ is more
intimate with node $t$ compared to node $i$ after the sufficient
evolution steps (Sup. Theorem 2). This means that the message will
not pass a node twice in the one sending process and always can
reach the target $t$.
\begin{figure}
\center
\includegraphics[width=8cm]{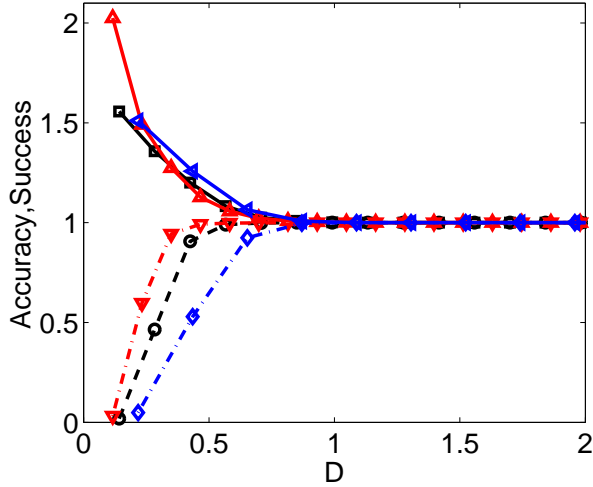}
\caption{Orientation ability. The numerical experiments is done in
the in the same networks as in Fig.1. The horizontal axis $D$ means
the evolving time for intimacy degree is $D\times AL$ ($AL$ is the
average shortest path length). Accuracy is defined as the value of
$\frac{L}{L_{s}}$, where $L_{s}$, $L$ denote the total length of the
pathes of successful searches and the corresponding total length of
shortest path length respectively. Success is defined as the
$\frac{S_{d}}{S}$, where $S_{d}$, $S$ denote the total number of
success searches and the corresponding total number of searches
respectively. If within the searching steps of $2\times AL$ in each
network, the message has not reached the target, the searching is
defined as a failure search. From the plot we can conclude that when
the evolved time is comparative to the average length of the
shortest path, Orientation will work very well.}\label{Orientation
ability}
\end{figure}

In order to demonstrate our model works well in finding short
chains, we define accuracy and success to evaluate the performance
of our algorithm. Here accuracy $L_s/L_{search}$ means the
consistence when the paths searched from algorithm is compared with
the shortest paths, and success is the rate of success searches to
reach the target in a given time steps, where, $L_s$ is the sum of
total shortest path lengths and $L_{search}$ is the corresponding
sum of total searched path lengths. We do the numerical experiments
on following three types artificial network : WS network which was
presented by Watts and Strogatz \cite{first issue b}, K network
which was presented by Kleinberg \cite{navigation brief nature} and
BA network which was presented by Barabasi and Albert \cite{BA
model}. All artificial networks have 1000 nodes and the average
degree is 3. Specially, the K network is based on 1-dimensional
lattice, each node connects the two nearest neighbors and has one
long-rang connection with clustering exponent $\alpha=1$. The
numerical results indicate our algorithm works well (as shown in
Fig.\ref{Orientation ability}).

The time complexity for the evolution of intimacy vectors is
$O(log(n)n^{2})$. When intimacy vectors are established, the
expectation of the time complexity for search is only $O(log(n))$ in
small world networks.

\section{Summary}

In summary, the intimacy based orientation can well explain the
small world phenomena. It shows that the individuals can search the
short path in social networks with only the local information. The
basic reason is that the intimacy degree of the individuals will
decay with the length of the shortest path exponentially. Moreover,
it also can explain why successful social search is conducted
primarily through intermediate to weak strength ties, does not
require highly connected ``hubs" to succeed. This phenomenon was
observed in real-world experiment \cite{new experiment} and cannot
be explained by the previous researches. For application, the space
complexity (for each node, we need an $n$-dimensional vector) of
Orientation is higher than the previous algorithm such as Navigation
\cite{navigation full}. But it is a decentralized algorithm
naturally and the space complicity is not a challenging problem in
decentralized computing system. We can also use community structures
in networks to reduce the space complexity. Our orientation method
has potential applications in P2P search system, traffic navigation
system, Internet routing and so on in the future.

\section*{Acknowledgement}
This work is partially supported by 985 Projet and NSFC under the
grant No.$60534080$, No.$70771011$ and No.$70471080$.

\section{Supplementary}

\subsection{Theorems and proof}
In a network, if there exist a path form node $p_{m}$ to node
$p_{0}$: $p_{m}\rightarrow p_{m-1}\rightarrow\cdots\rightarrow
p_{0}$, then there must be a corresponding intimacy spreading path
$p_{0}\rightarrow p_{1}\rightarrow\cdots\rightarrow p_{m}$ which we
call intimacy spreading path form $p_{0}$ to $p_{m}$. That's to say
we send message alone  $p_{m}\rightarrow
p_{m-1}\rightarrow\cdots\rightarrow p_{0}$, and get nodes
information alone $p_{0}\rightarrow
p_{1}\rightarrow\cdots\rightarrow p_{m}$. It is obvious that there
exist a intimacy spreading path from $p_{0}$ to $p_{m}$, there must
be a path from $p_{m}$ to $p_{0}$.

\textbf{Theorem 1}: Suppose $a>>1$ and there exist a constant $H$
such that for any positive $m$,
$\prod_{i=1}^{m}(k_{i}a+k_{i}-1)\approx H^{m}$ in statistic sense,
where $k_{i}$ is the out degree of node $i$ (it is strict in lattice
based networks). Then intimacy degree of $p_{m}$ possessed about
$p_{0}$ will decay with the length of their shortest path
exponentially, and will decay inversely to the degree of the node by
which intimacy passes.

\textbf{Proof}:

Suppose $p_{i}$ has $k_{i}$ neighbors,
$N_{p_{1}}^{1},N_{p_{1}}^{2},\cdots,N_{p_{1}}^{k_{1}}$ denote all
the neighbors of $p_{1}$ and $p_{0}$ is a neighbor of $p_{1}$,
$v^{T}_{x}(y)$ denotes the intimacy degree of node $x$ to node $y$
after $T$ steps evolution, $v^{0}_{x}(y)=a$ if $y$=$x$, otherwise
$0$. Without losing generality we let $p_{0}=N_{p_{1}}^{1}$.

$\because\ $ $p_{0}\rightarrow p_{1}\rightarrow\cdots\rightarrow
p_{m}$ is an intimacy spreading path from $p_{0}$ to $p_{m}$.

$\therefore\ $ from the Eq.\ref {evolution} we have
$v^{T}_{p_{1}}(p_{0})=\frac{\sum_{i=1}^{k_{1}}v^{T-1}_{N_{p_{1}}^{i}}(p_{0})}{\sum_{i=1}^{k_{1}}\sum_{j=1}^{n}v^{T-1}_{N_{p_{1}}^{i}}(j)-\sum_{i=1}^{k_{1}}v^{T-1}_{N_{p_{1}}^{i}}(p_{1})}
=\frac{a+\sum_{i=2}^{k_{1}}v^{T-1}_{N_{p_{1}}^{i}}(p_{0})}{k_{1}a+k_{1}-\sum_{i=1}^{k_{1}}v^{T-1}_{N_{p_{1}}^{i}}(p_{1})}$

$\because\ 0<v^{T-1}_{N_{p_{1}}^{i}}(p_{0})\leq 1, i=1,\cdots,k_1$

$\therefore\ \frac{a}{k_{1}a+k_{1}}\leq
v^{T-1}_{p_{1}}(p_{0})\leq\frac{a+k_{1}}{k_{1}a}$

$\because\ a\gg1$

$\therefore\ v^{T}_{p_{1}}(p_{0})\thickapprox\frac{1}{k_{1}}$,
$T=1,2\cdots +\infty$

$\therefore\
v^{T}_{N_{p_{i}}^{j}}(p_{i})\thickapprox\frac{1}{k_{i}}$,
$T=1,2\cdots +\infty$

\begin{figure}
\center
\includegraphics[width=7cm]{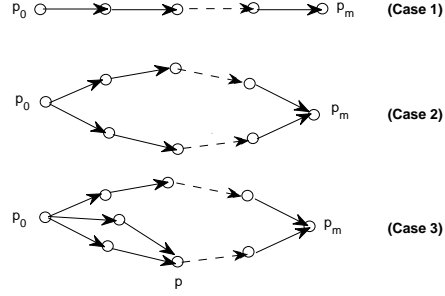}
\caption{The plot represents three cases in the proof. Case 1
denotes there are only one intimacy spreading shortest path from
$p_0$ to $p_m$. Case 2 shows there are two dependent shortest pathes
and case 3 denotes 3 independent shortest pathes in which two pathes
encounter at node $p$.}\label{proof_plot}
\end{figure}

Case 1 (see Fig. \ref{proof_plot} case 1):\ suppose there exist only
one shortest path ($p_{m}\rightarrow
p_{m-1}\rightarrow\cdots\rightarrow p_{0}$) from $p_{m}$ to $p_{0}$,
then when the intimacy of $p_{0}$ spread to $p_{i}$ currently alone
$p_{0}\rightarrow p_{1}\rightarrow\cdots\rightarrow p_{m}$.

Then the first intimacy degree (It means the intimacy degree
possessed in the first time of one pair of nodes)
\,\,\,$v_{p_{i}}(p_{0})=\frac{v_{p_{i-1}}(p_{0})}{k_{i}a+k_{i}-\sum_{h=1}^{k_{i}}v_{N_{p_{i}}^{h}}(p_{i})}$
$\approx\,\frac{v_{p_{i-1}}(p_{0})}{k_{i}a+k_{i}-k_{i}\frac{1}{k_{i}}}=\frac{v_{p_{i-1}}(p_{0})}{k_{i}a+k_{i}-1}$
where, $N_{p_{i}}^{1},\cdots, N_{p_{i}}^{k_{i}}$ are $k_{i}$
neighbors of $p_{i}$ and $N_{p_{i}}^{1}=p_{i-1}$. (The above
equation also implies the intimacy degree will decay inversely
($\frac{1}{k_{i}a}$) to the degree of the node by which intimacy
passes.)

In this way we
have:\,\,\,$v_{p_{m}}(p_{0})=a\prod_{i=1}^{m}\frac{1}{k_{i}a+k_{i}-1}$

$\because\,\,\, \prod_{i=1}^{m}(k_{i}a+k_{i}-1)\approx H^{m}$

Then we have \,\,\,$v_{p_{m}}(p_{0})\approx aH^{-m}$

Case 2 (see Fig. \ref{proof_plot} case 2):\ When there are $r$
independent (means there is no common node for each pair of shortest
pathes) shortest intimacy spreading pathes from $p_{0}$ to $p_{m}$
through $r$ neighbors of $p_{m}$. Assume the length of the shortest
intimacy spreading pathes from $p_{0}$ to $p_{m}$ is $m$,then we can
easily get that

$v_{p_{m}}(p_{0})\approx\frac{raH^{-(m-1)}}{ak_{m}+k_{m}-1}\,\approx\,raH^{-m}$

Case 3 (see Fig. \ref{proof_plot} case 3):\ When there are $r$
shortest intimacy spreading pathes from $p_{0}$ to $p_{m}$ in which
$\beta$ shortest intimacy spreading pathes are dependent and other
$r-\beta$ are independent. All the situations are equal to this
situation: all of the $\beta$ dependent intimacy spreading pathes
encounter at node $p$ and they are independent from $p_{0}$ to $p$
and there only one shortest intimacy spreading path for $p$ to
$p_{m}$. Assume the length of shortest intimacy spreading pathes
from $p_{0}$ to $p_{m}$ is $m$, from $p_{0}$ to $p$ is $m_{1}$ and
$p$ to $p_{m}$ is $m_{2}=m-m_{1}$.

According to case $1$ and case $2$ we have

$v_{p_{m}}(p_{0})\approx(r-\beta)aH^{-m}+(taH^{-m_{1}})H^{-m_{2}}\,=\,raH^{-m}$

Now, the task we face is to prove $v^{T}_{p_{m}}(p_{0})\approx
raH^{-m}$ for any $T>=m$.

Obviously, $v^{T}_{p_{m}}(p_{0})$ equal the sum of the each step
($m$ to $T$) first intimacy degree, then we have :

$\,raH^{-m}=v_{p_{m}}(p_{0})\leq v^{T}_{p_{m}}(p_{0})\leq
\,raH^{-m}+a[\frac{K^{m+1}}{a^{m+1}}+\frac{K^{m+2}}{a^{m+2}}+,\cdots,+\frac{K^{T}}{a^{T}}]$
where $K$ is the maximum out degree of all nodes.

Thus, $v^{T}_{p_{m}}(p_{0})\leq
raH^{-m}+a[\frac{K^{m+1}}{a^{m+1}}+\int_{m+1}^{+\infty}(\frac{K}{a})^xdx]\leq
raH^{-m}+2a\frac{K^{m+1}}{a^{m+1}}$

$\because\,\,\,\lim_{a\rightarrow\infty}\frac{2\frac{K^{m+1}}{a^{m+1}}}{H^{-m}}=0$

$\therefore\,\,\,$ for sufficient large $a$,
$v^{T}_{p_{m}}(p_{0})\approx raH^{-m}$.

\textbf{Theorem 2}: For a connected network and $a>1$. If $i$ is not
the target $t$ then there must be at least one neighbor $q$ of node
$i$, $q$ is more intimate with target $t$ compared to node $i$ after
sufficient long time evolution. This means that the message will not
pass a node twice in the one sending process and always can reach
the target $t$.

\textbf{Proof}:

$\because$ the network is connected

$\therefore$ node $i$ at lest has a neighbor.

Assume $i$ has $k_{i}$ neighbors which are
$N_{i}^{1},N_{i}^{2},\cdots,N_{i}^{k_{i}}$.

$\because\
v^{+\infty}_{i}(j)=\frac{v^{+\infty}_{N_{i}^{1}}(j)+v^{+\infty}_{N_{i}^{2}}(j)+\cdots+v^{+\infty}_{N_{i}^{k_{i}}}(j)}{k_{i}a+k_{i}-\sum_{d=1}^{k_{i}}v^{+\infty}_{N_{i}^{d}}(i)}$
and\
$k_{i}a+k_{i}-\sum_{d=1}^{k_{i}}v^{+\infty}_{N_{i}^{d}}(i)>k_{i}$

$\therefore\,\,\,v^{+\infty}_{i}(j)<\max\{v^{+\infty}_{N_{i}^{1}}(j),
v^{+\infty}_{N_{i}^{2}}(j), \cdots,
v^{+\infty}_{N_{i}^{k_{i}}}(j)\}$

Therefore the above $q$ must be existed and in the a sending
process, the current message holder will always more intimacy with
target $t$ than the all previous message holder which implies that
the message will not pass a node twice. Because the number of node
of a network is finite, the message will and can reach the target
$t$.


\begin{thebibliography}{99}
\bibitem{the oldest experiment}Milgram, S. Psych. Today 2, 60 (May 1967).
\bibitem{Earlier}Karinthy, F., Chains. Everything is different. Budapest, (1929).
\bibitem{Play} Guare, J. Six degrees of separation: a play. Vintage, New York,
(1990).
\bibitem{new experiment}Dodds, P. S., Muhamad,R., Watts, D. J. An Experimental Study of Search in Global Social Networks Science \textbf{301}, 827 (2003)
\bibitem{News}Fiona Macrae, Microsoft proves you ARE just six degrees of separation from anyone in the
world, Mailonline, Science and Tech, 04th August 2008,
http://www.dailymail.co.uk/sciencetech/article-1041077/Microsoft-proves-ARE-just-degrees-separation-world.html.
\bibitem{first issue b}Watts, D., Strogatz, S. Collective dynamics of small world Networks. Nature \textbf{393}, 440-442 (1998).
\bibitem{SWNature}S. H. Strogatz, Nature \textbf{410}, 268 (2001).
\bibitem{BA model}Albert, R. and Barab$\acute{a}$si, A.-L. Statistical mechanics of complex networks. Reviews of Modern Physics, \textbf{74}, (2002).
\bibitem{Newman-review} Newman, M. E. J. The structure and function of complex networks. SIAM Review, \textbf{45}, 167¨C256, (2003).
\bibitem{navigation brief nature}Kleinberg, J. M.  Navigation in a small world. Nature \textbf{406}, 845
(2000).
\bibitem{navigation full}Kleinberg, J. M. The Small-World Phenomenon: An Algorithmic Perspective. In Proceedings of the 32nd ACM Symposium on Theory of Computing, (2000).
\bibitem{Kleinberg hierarchical model}Kleinberg, J. M. Small-World Phenomena and the Dynamics of Inforamtion. In Advances in Neural Information Processing Systems (NIPS) 14, (2001).
\bibitem{full model navigation}Watts, D. J.,  Dodds,P. S. and Newman, M. E. J. Identity and search in social networks. Science, \textbf{296}, 1302-1305, (2002).
\bibitem{power-law networks search} Adamic, L., Lukose, R.  A., Puniyani, A. R. and  Huberman, B. A. Search in power-law networks. Physical Review E, \textbf{64}, 46135, (2001).
\bibitem{Use Kleinberg search}Liben-Nowell, D.,  Novak, J.,  Kumar, R.,  Raghavan, P. and Tomkins, A. Geograph routing in social networks. In Proceedings of the National Academy of Science, \textbf{102}, 11623-11628, (2005).
\bibitem{Fraigniauda}Fraigniaud, P., Gavoille, C., Kosowski, A., Lebhar, E. and Lotker, Z. in Proc. Nineteenth Annual ACM Symp. on Parallel Algorithms and Architectures. 1-7, ACM, (2007).
\bibitem{Fraigniaudb}Fraigniaud, P. and Gavoille, C. in Proc. Twentieth Annual Symposium on Parallelism in Algorithms and Architectures. 62-69, ACM, (2008) SPAA. 29.
\bibitem{Oskar licentiate thesis}Sandberg, O. Searching in a Small World, University of Gothenburg and Chalmers Technical University (licentiate thesis), (2005).
\bibitem{Analyzing kleinberg}Martel, C. and Nguyen, V. Analyzing kleinberg¡¯s (and other) small-world models. In PODC ¡¯04: Proceedings of the twentythird annual ACM symposium on Principles of distributed computing, pages 179¨C188, New York, NY, USA, ACM Press. (2004).
\bibitem{one-d-Analyzing kleinberg}Barriere, L., Fraigniaud, P., Kranakis, E., Krizanc, D. Efficient Routing in Networks with Long Range Contacts, Proc. 15th Intr. Symp. on Dist. Computing, DISC 01, 270-284, (2001).
\bibitem{power-law networks Kleinberg search}Adamic, L. and Adar, E. How to search a social network. Social Networks, \textbf{27}, 187-203, (2005).
\bibitem{scale free metric navigation}Bogu$\tilde{n}\acute{a}$, M., Krioukov, D. and Claffy, K. C. Navigability of complex networks. Nature Physics, advance online publication, NPHYS1130,
(2008).

\end{thebibliography}
\end{document}